\documentclass[
onlinecite, reprint, 
superscriptaddress,
nobalancelastpage,
amsmath,amssymb,
aps,prl
]{revtex4-2}

\usepackage{makecell}
\usepackage{graphicx}
\usepackage{dcolumn}
\usepackage{bm}
\usepackage{longtable}
\usepackage{physics}
\usepackage{amsmath}
\usepackage{multirow}
\usepackage[table,dvipsnames]{xcolor}
\usepackage{colortbl}
\usepackage{booktabs}
\usepackage{soul}
\usepackage{comment}
\usepackage[
    colorlinks=true,
    linkcolor=blue,
    citecolor=blue,
    urlcolor=blue
]{hyperref}

\begin{document}
\title{Zeeman Shift and Injection Currents under Light Irradiation}

\author{Ruizhi Dong}
\affiliation{Key Lab of advanced optoelectronic quantum architecture and measurement (MOE),
Beijing Key Laboratory of Quantum Matter State Control and Ultra-Precision Measurement Technology,
and School of Physics, Beijing Institute of Technology, Beijing 100081, China}

\author{Yihua Xiao}
\affiliation{Key Lab of advanced optoelectronic quantum architecture and measurement (MOE),
Beijing Key Laboratory of Quantum Matter State Control and Ultra-Precision Measurement Technology,
and School of Physics, Beijing Institute of Technology, Beijing 100081, China}

\author{Pengfei Wang}
\affiliation{Key Lab of advanced optoelectronic quantum architecture and measurement (MOE),
Beijing Key Laboratory of Quantum Matter State Control and Ultra-Precision Measurement Technology,
and School of Physics, Beijing Institute of Technology, Beijing 100081, China}

\author{Ruixiang Fei}
\email{rfei@bit.edu.cn}
\affiliation{Key Lab of advanced optoelectronic quantum architecture and measurement (MOE),
Beijing Key Laboratory of Quantum Matter State Control and Ultra-Precision Measurement Technology,
and School of Physics, Beijing Institute of Technology, Beijing 100081, China}

\begin{abstract}
The conversion of light into electrical currents enables energy harvesting and photodetection. Conventional second-order dc photocurrents are driven by the electric component of light through electric-dipole coupling and require broken inversion symmetry. Here, we show that the electric and magnetic components of light jointly generate Zeeman shift and injection currents.
Using quantum perturbation theory, we derive explicit microscopic
expressions for their conductivity tensors.
We systematically classify the symmetry constraints on these responses
and show how they differ from those governing conventional photocurrents.
In particular, both responses can occur in centrosymmetric systems
with spin--orbit coupling or noncollinear magnetic order,
irrespective of time-reversal symmetry.
A centrosymmetric kagome model exhibits broadband responses.
First-principles calculations for Mn$_3$Sn further predict Zeeman
injection and shift current densities of
$10^{2}$--$10^{3}~\mathrm{A/m^{2}}$ and
$1$--$10~\mathrm{A/m^{2}}$, respectively, under illumination with an electric-field amplitude
$E=10^{4}~\mathrm{V/m}$.
These magnitudes place both currents within experimental reach.
\end{abstract}
\maketitle

\textbf{Introduction}---
Photocurrent generation links microscopic 
optical excitations to macroscopic transport,
enabling energy conversion and optoelectronic applications~\cite{Choi2009,Grinberg2013,Rosencher2002}. 
Conventional theories of photocurrent generation focus on electric-dipole coupling to the optical electric field. Within this framework, the response is closely linked to the quantum geometry of Bloch states, including Berry connections and curvature~\cite{Morimoto2016,Moore2010,deJuan2017,Ma2017,Ji2019}, as well as the Riemannian geometry of interband optical transitions~\cite{Ahn2022,Fregoso2019}.
Beyond wave-function geometry, photocurrents can reveal geometric features of band dispersions encoded in energy eigenvalues~\cite{Fregoso2018,Dong2026,Sarkar2025}.
These connections provide a geometric perspective on charge and spin
photocurrent generation~\cite{Ahn2022,Dong2026} and help elucidate
the symmetry constraints and polarization selection rules governing
these responses~\cite{Young2013,Fei2021,Xu2021,Ahn2020,Watanabe2021}.

A prominent example is the bulk photovoltaic effect, which offers a route to junction-free photodetection and solar-energy conversion~\cite{sipe2000,Glass1974,vonBaltz1981,Young2012,Osterhoudt2019}.
Its shift-current contribution is governed by a gauge-invariant
shift vector that characterizes the real-space displacement
associated with an interband optical transition~\cite{sipe2000,Morimoto2016}.
The optical magnetic field, however, also couples to electron spin through the Zeeman interaction. Its contribution to photocurrent generation is usually neglected because it is presumed small~\cite{Junck2013}, yet whether this coupling can give rise to qualitatively new mechanisms remains largely unexplored.


In this work, we develop a microscopic theory of photocurrents arising from interference between electric-dipole and optical Zeeman transition amplitudes. We identify two additional contributions, which we term Zeeman shift and injection currents. 
We systematically classify these responses and establish how their symmetry properties differ from those of conventional electric-dipole shift and injection currents.
Combining model analyses with first-principles calculations of candidate materials, we further predict that both mechanisms yield experimentally detectable photocurrents.

\textbf{Phenomenological description of Zeeman shift and injection currents}---
To elucidate the physical origin of these currents, we first consider a Hamiltonian incorporating both the conventional 
electric-dipole interaction~\cite{sipe2000}
and the omitted optical spin Zeeman coupling:
\begin{align}
\label{eq_ham}
\hat{H}
= \hat{H}_0 - e\mathbf{E}(t)\cdot\hat{\mathbf{r}}
- \frac{g}{2}\mu_B\mathbf{B}(t)\cdot\hat{\boldsymbol{\sigma}},
\end{align}
where $\hat{H}_0$ is the unperturbed Bloch Hamiltonian, and
$\mathbf{E}(t)$ and $\mathbf{B}(t)$ denote the electric and magnetic
fields of light, respectively. Here, $g$ is the electron $g$ factor,
$\mu_B$ is the Bohr magneton, and $\hat{\boldsymbol{\sigma}}$
is the vector of Pauli matrices.


The second term in Eq.~\ref{eq_ham} describes the well-studied electric-dipole coupling, which generates resonant 
dc photocurrents at second order in the optical field~\cite{sipe2000}. 
The shift and injection contributions can be expressed schematically using Fermi's golden rule as
\begin{align}
J_{\mathrm{shift}} \propto \mathcal{W}_{n\rightarrow m}^{EE} R_{mn},\qquad
J_{\mathrm{inj}} \propto \mathcal{W}_{n\rightarrow m}^{EE} \Delta_{mn} \tau
\label{eq_Fermi_EE}
\end{align}
where $R_{mn}$ is the shift vector describing the real-space displacement associated with the optical transition, $\Delta_{mn}=v_m-v_n$ is the corresponding change in band velocity, and $\tau$ is the relaxation time governing the steady-state injection current.
The electric-dipole transition rate entering
both contributions is given by
\begin{align}
\mathcal{W}_{n\rightarrow m}^{EE}
&=\frac{2\pi}{\hbar^2}|M_{mn}^{E}(\omega)|^2
\delta(\omega_{mn}-\omega) \nonumber\\
&=\frac{2\pi e^2}{\hbar^2}
r_{mn}^{b}r_{nm}^{c}
E_{\omega}^{b}E_{-\omega}^{c}
\delta(\omega_{mn}-\omega),
\label{eq_EE_ratio}
\end{align}
Here, $M_{mn}^{E}(\omega)=-eE_{\omega}^{b}r_{mn}^{b}$ denotes the electric-dipole transition amplitude, $r_{mn}^{b}$ the interband position matrix element, and $\omega_{mn}=(\varepsilon_m-\varepsilon_n)/\hbar$ the transition frequency.
Its transition weight is determined by $Q_{mn}^{bc}=r_{mn}^{b}r_{nm}^{c}$,
the quantum geometry of the Bloch states.

\begin{figure}[!t]
\includegraphics[width=8.5cm]{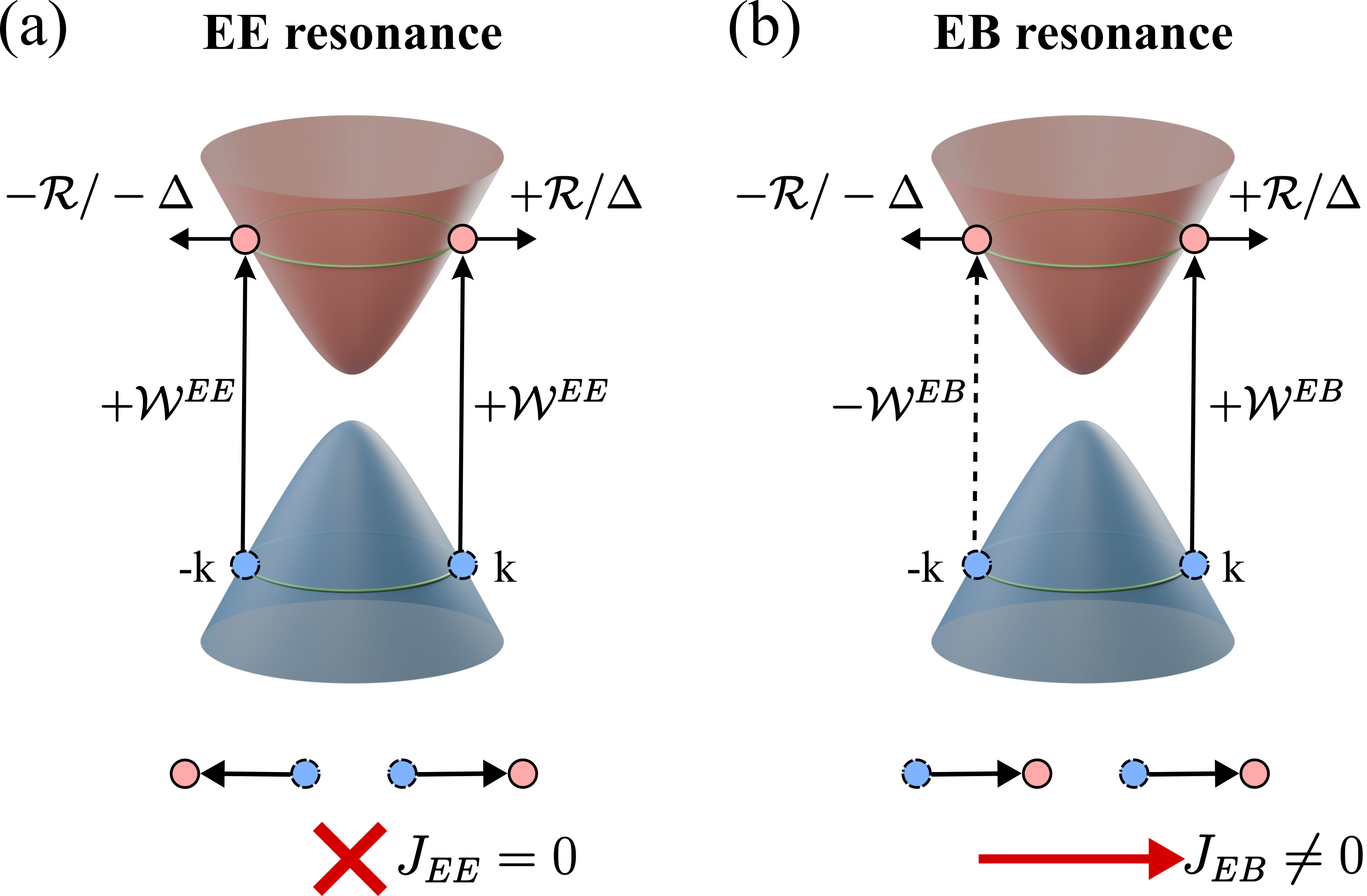}
\caption{
Fermi's golden rule picture of photocurrent generation in a centrosymmetric system.
(a) For electric-dipole ($EE$) transitions, the inversion-even rate $\mathcal{W}^{EE}$ and inversion-odd shift vector $\mathcal{R}$ or velocity difference $\Delta$ yield opposite contributions at $\pm\mathbf{k}$, giving $J_{EE}=0$.
(b) The electric-dipole--Zeeman ($EB$) interference term is inversion odd, $\mathcal{W}^{EB}(\mathbf{k})=-\mathcal{W}^{EB}(-\mathbf{k})$. The corresponding photocurrent contributions at $\pm\mathbf{k}$ therefore add, allowing a finite $J_{EB}$.
}
\label{Fig1}
\end{figure}

With the optical Zeeman coupling included, first-order time-dependent perturbation theory yields the total transition amplitude for $n\rightarrow m$ as
\begin{align}
M_{mn}(\omega)
=-eE_{\omega}^{b}r_{mn}^{b}
-\frac{g}{2}\mu_B B_{\omega}^{\beta}\sigma_{mn}^{\beta},
\end{align}
Repeated Cartesian indices $b,\beta=x,y,z$ are summed. The corresponding transition rate follows from Fermi's golden rule:
\begin{align}
\mathcal{W}_{n\rightarrow m}
&=\frac{2\pi}{\hbar^2}|M_{mn}(\omega)|^2
\delta(\omega_{mn}-\omega)\notag\\
&=\mathcal{W}_{n\rightarrow m}^{EE}
+\mathcal{W}_{n\rightarrow m}^{EB}
+\mathcal{W}_{n\rightarrow m}^{BB},
\end{align}
The $EE$ and $BB$ terms arise solely from the electric-dipole and spin Zeeman couplings, respectively, whereas $EB$ captures interference between their transition amplitudes (see Supplemental Material~\cite{supp} for details).
The $BB$ contribution transforms identically to $EE$ and is quadratic in the optical Zeeman coupling.
We therefore neglect it in the weak-coupling limit and retain the $EB$ interference term, which is linear in the Zeeman coupling,
\begin{align}
\mathcal{W}_{n\rightarrow m}^{EB}
=
\frac{2\pi e g\mu_B}{\hbar^2}
\mathrm{Re}
\left[
r_{mn}^{b}\sigma_{nm}^{\beta}
E_{\omega}^{b}B_{-\omega}^{\beta}
\right]
\delta(\omega_{mn}-\omega).
\end{align}

Here, $\sigma_{nm}^{\beta}=\langle u_{n\mathbf{k}}|\hat{\sigma}^{\beta}|u_{m\mathbf{k}}\rangle$ denotes the interband matrix element of the Pauli matrix $\hat{\sigma}^{\beta}$. The transition factor $\mathcal{Z}_{mn}^{b\beta}=r_{mn}^{b}\sigma_{nm}^{\beta}$ is 
associated with the recently proposed Zeeman quantum geometry~\cite{Xiang2025}.
Previous studies connected this geometry and related structures to linear responses, including gyrotropic currents and field-induced magnetization~\cite{Xiang2025,Dong2020,jia2026}.
Here, it enters as a geometric factor in the $EB$-resonant photocurrent,
which is physically distinct from these previously studied responses.

By analogy with conventional $EE$ shift and injection currents, we term the corresponding $EB$ contributions Zeeman shift and injection currents. Schematically,
\begin{align}
J_{\mathrm{Zshift}} \propto \mathcal{W}_{n\rightarrow m}^{EB} R_{mn},\qquad
J_{\mathrm{Zinj}} \propto \mathcal{W}_{n\rightarrow m}^{EB} \Delta_{mn}\tau.
\label{eq_Fermi_Zeeman}
\end{align}

The $EB$ contribution obeys symmetry constraints fundamentally
distinct from those of the conventional $EE$ response.

Spatial inversion $\mathcal{P}$ provides a particularly clear illustration of this distinction. In centrosymmetric systems, 
$r_{mn}^{b}r_{nm}^{c}$ is $\mathcal{P}$-even, whereas $R_{mn}$  and $\Delta_{mn}$ are $\mathcal{P}$-odd. The $EE$ photocurrent contributions at $\mathbf{k}$ and $-\mathbf{k}$ therefore cancel according to Eq.~\ref{eq_Fermi_EE}, as illustrated in Fig.~\ref{Fig1}(a). By contrast, 
$\mathcal{Z}_{mn}^{b\beta}=r_{mn}^{b}\sigma_{nm}^{\beta}$ 
is $\mathcal{P}$-odd, making its product with either $R_{mn}$  or $\Delta_{mn}$ $\mathcal{P}$-even. The $EB$ contributions from opposite momenta thus add according to Eq.~\ref{eq_Fermi_Zeeman}, allowing a finite photocurrent in centrosymmetric systems, as shown in Fig.~\ref{Fig1}(b).

\textbf{Perturbative derivation of Zeeman shift and injection currents}---
The transition-rate picture captures the physical origin of the $EB$
response. A microscopic treatment is needed to determine the full
conductivity tensors and their geometric structure. We therefore derive
explicit expressions for the Zeeman shift and injection currents using
density-matrix perturbation theory.

Within the density-matrix formalism, the current density along direction $a$ is obtained from the expectation value of the current-density operator:
\begin{align}
\label{current_density}
J^a
= \frac{e}{V}\mathrm{Tr}\!\left[\hat{v}^a\rho\right]
= \mathrm{Tr}\!\left[\hat{J}^a\rho\right].
\end{align}

Guided by Sipe \textit{et al.}'s treatment of $EE$ shift and injection currents~\cite{sipe2000}, we express the dc response entirely in terms of the intraband current operator,
\begin{equation}
\label{eq:intraband_current}
\hat{J}_{\mathrm{intra}}^{a}
=\hat{J}^{a}-\hat{J}_{\mathrm{inter}}^{a}
=\hat{J}^{a}-\frac{d}{dt}\hat{P}_{\mathrm{e}}^{a},
\end{equation}
where $\hat{P}_{\mathrm{e}}^{a}$ is the interband polarization operator. Although the interband current has no dc component~\cite{supp}, evaluating it through $\hat{P}_{\mathrm{e}}^{a}$ and subtracting it from the total current is convenient in the second-quantized formalism. This approach facilitates higher-order perturbative calculations and the separation of shift- and injection-current contributions~\cite{sipe2000,supp}.

Since $\hat{P}_{\mathrm{e}}^{a}$ has no explicit time dependence,
the Heisenberg equation gives
$d\hat{P}_{\mathrm{e}}^{a}/dt=(i\hbar)^{-1}
[\hat{P}_{\mathrm{e}}^{a},\hat{H}]$.
Substituting the second-quantized form of the Hamiltonian in
Eq.~\ref{eq_ham} into this equation yields
\begin{equation}
\label{eq:interband_polarization_eom}
\begin{aligned}
\frac{d\hat{P}_{\mathrm{e}}^{a}}{dt}
&=\frac{1}{i\hbar}
[\hat{P}_{\mathrm{e}}^{a},\hat{H}_{0}]
-\frac{V}{i\hbar}
[\hat{P}_{\mathrm{e}}^{a},\hat{P}_{\mathrm{e}}^{b}+\hat{P}_{\mathrm{i}}^{b}]E^{b}(t) \\
&\quad-\frac{V}{i\hbar}
[\hat{P}_{\mathrm{e}}^{a},\hat{M}^{\beta}]B^{\beta}(t).
\end{aligned}
\end{equation}
where $\hat{P}_{\mathrm{i}}^{b}$ denotes the intraband polarization operator and $\hat{M}^{\beta}=(g\mu_B/2V)\sum_{n,m,\mathbf{k}}\sigma_{nm}^{\beta}(\mathbf{k})a_{n\mathbf{k}}^{\dagger}a_{m\mathbf{k}}$ is the spin magnetization operator. 
Here, $a_{n\mathbf{k}}^{\dagger}$ ($a_{n\mathbf{k}}$) creates (annihilates) an electron in the Bloch state $(n,\mathbf{k})$.
The last commutator arises from the optical Zeeman coupling.


Evaluating the commutators yields the intraband current operator (see Supplemental Material~\cite{supp} for the derivation):
\begin{align}
\hat{J}_{\mathrm{intra}}^a
={}&\frac{e}{V}\sum_{nm\mathbf{k}}
\Bigl[
\Bigl(v_{nn}^a
-\frac{e}{\hbar}
(\mathbf{E}\times\boldsymbol{\Omega}_n)^a\Bigr)
\delta_{nm}
\notag\\
&\quad
-\frac{e}{\hbar}E^b(t) r_{nm;a}^{\,b}
-\frac{g\mu_B}{2\hbar}B^\beta(t)\sigma_{nm;a}^{\,\beta}
\Bigr]
a_{n\mathbf{k}}^\dagger a_{m\mathbf{k}}.
\label{current_oper}
\end{align}
Here, $\boldsymbol{\Omega}_n$ is the Berry curvature of band $n$, and
$\sigma_{nm;a}^{\,\beta}
=\partial_{k_a}\sigma_{nm}^{\beta}
-i(A_{nn}^{a}-A_{mm}^{a})\sigma_{nm}^{\beta}$
is the covariant derivative of the spin matrix element, with
$A_{nn}^{a}=i\langle u_{n\mathbf{k}}|
\partial_{k_a}u_{n\mathbf{k}}\rangle$
the Berry connection.
The final term, absent in the electric-dipole formulation, reflects optical Zeeman coupling.


To evaluate the current expectation value using
Eq.~\ref{current_density}, we next solve the Liouville--von Neumann
equation for the density matrix,
\begin{equation}
\label{eq:liouville}
i\hbar\frac{\partial\rho}{\partial t}=[\hat{H},\rho],
\end{equation}
where $\hat{H}$ is the full Hamiltonian in Eq.~\ref{eq_ham},
including both electric-dipole and Zeeman couplings.
Expanding the density matrix perturbatively in powers of the electric
and magnetic fields as
$\rho=\rho^{(0)}+\rho^{(1)}+\rho^{(2)}+\cdots$,
the first-order correction is
$\rho_{mn}^{(1)}(t)
=
\bar\rho_{mn}^{(1),b}
E_{\omega_b}^{b}e^{-i\omega_b t}
+
\bar\rho_{mn}^{(1),\beta}
B_{\omega_\beta}^{\beta}e^{-i\omega_\beta t}$,
with
\begin{align}
\bar\rho_{mn}^{(1),b}
&=
\frac{ie}
{\hbar(\omega_{mn}-\omega_b-i\eta)}
\left[
\delta_{mn}\partial_{k_b}f_n
+i r_{mn}^b f_{mn}
\right],
\notag
\\
\bar\rho_{mn}^{(1),\beta}
&=
-
\frac{
g\mu_B f_{mn}\sigma_{mn}^\beta
}
{2\hbar(\omega_{mn}-\omega_\beta-i\eta)}.
\label{eq:rho1B}
\end{align}
where $\eta=1/\tau$ is the phenomenological relaxation rate.
At second order, the density matrix 
contains $EE$, $EB$, and $BB$ contributions.
We focus on the 
$EB$ contribution,
$\rho_{mn}^{(2),EB}(t)
=
\bar\rho_{mn}^{(2),b\beta}
E_{\omega_b}^{b}
B_{\omega_\beta}^{\beta}
e^{-i(\omega_b+\omega_\beta)t}$,
with 
\begin{align}
\bar\rho_{mn}^{(2),b\beta}
= &
\frac{ie}
{\hbar(
\omega_{mn}-\omega_b-\omega_\beta-i\eta)}
\Bigg[
\bar\rho_{mn;b}^{(1),\beta}
\notag\\
&+
i\sum_l
\left(
\bar\rho_{ml}^{(1),\beta}r_{ln}^b
-
r_{ml}^b\bar\rho_{ln}^{(1),\beta}
\right)
\notag\\
&-
i\frac{g\mu_B}{2e}
\sum_l
\left(
\sigma_{ml}^\beta\bar\rho_{ln}^{(1),b}
-
\bar\rho_{ml}^{(1),b}\sigma_{ln}^\beta
\right)
\Bigg].
\label{eq:rho2_EB}
\end{align}

Combining the intraband current operator in Eq.~\ref{current_oper} with the density-matrix responses yields our central result for the resonant $EB$ dc photocurrent:
\begin{align}
\langle J_{EB}^{a}(t)\rangle
=&\frac{e}{V}\sum_{nm\mathbf{k}}
\bigg[
-\frac{e}{\hbar}r_{nm;a}^{b}\bar\rho_{mn}^{(1),\beta}
-\frac{g\mu_B}{2\hbar}\sigma_{nm;a}^{\beta}\bar\rho_{mn}^{(1),b}
\notag\\
&+v_{nn}^{a}\delta_{nm}\bar\rho_{mn}^{(2),b\beta}
\bigg]
E_{\omega_b}^{b}B_{\omega_\beta}^{\beta}
e^{-i(\omega_b+\omega_\beta)t}.
\label{eq:JEB}
\end{align}
For the dc response, $\omega_b+\omega_\beta=0$, implying
$\omega_b=-\omega_\beta=\pm\omega$.
The first two terms yield the \textit{Zeeman shift current},
while the last gives the \textit{Zeeman injection current},
both arising from resonant $EB$ interference.

We first consider the Zeeman shift current,
$\langle J_{\mathrm{Zshift}}^a\rangle
=
\eta_{\mathrm{Zshift}}^{a;b\beta}(0;-\omega,\omega)
E_{-\omega}^b B_{\omega}^{\beta}
+\mathrm{c.c.}$,
with the response tensor given by (see Supplemental Material~\cite{supp} for the detailed derivation)
\begin{align}
\label{Zshift_purt}
\eta^{a;b\beta}_\mathrm{Zshift}(0;-\omega,\omega)
=&
\frac{\pi g\mu_B e^2}{2\hbar^2}
\sum_{nm}\int_{\mathbf{k}}
r_{nm}^b\sigma_{mn}^\beta
\left(
R_{mn}^{\beta;a}-R_{nm}^{b;a}
\right)
\notag\\
&\cdot
f_{nm}
\delta(\omega_{mn}-\omega).
\end{align}
Here,
$R_{nm}^{b;a}
=i\partial_a\log r_{nm}^b+A_{nn}^a-A_{mm}^a$
is the conventional electric-dipole shift vector.
The mixed $EB$ response additionally involves a spin-dependent
shift vector,
$R_{mn}^{\beta;a}
=i\partial_a\log\sigma_{mn}^{\beta}+A_{mm}^a-A_{nn}^a$,
defined through the interband spin matrix element.
The microscopic perturbative derivation reveals this spin-dependent
shift as an additional contribution from Zeeman quantum geometry
beyond the schematic expression in Eq.~\ref{eq_Fermi_Zeeman}.

Turning to the injection contribution, we obtain
$\langle J_{\mathrm{Zinj}}^a\rangle=
\eta_{\mathrm{Zinj}}^{a;b\beta}(0;-\omega,\omega) E_{-\omega}^b B_{\omega}^{\beta}
+\mathrm{c.c.}$, with
\begin{align}
\label{Zinj_purt}
\eta_{\mathrm{Zinj}}^{a;b\beta}(0;-\omega,\omega)
=&
\frac{\pi g\mu_B e^2\tau}{\hbar^2}
\sum_{nm}\int_{\mathbf{k}}
r_{nm}^b\sigma_{mn}^\beta
\Delta_{mn}^a
\notag\\
&\cdot
f_{nm}\,
\delta(\omega_{mn}-\omega),
\end{align}
where $\Delta_{mn}^a=v_{mm}^a-v_{nn}^a$ is the group-velocity
difference between bands $m$ and $n$, and $\tau$ is the carrier lifetime.
This result reproduces the Fermi-golden-rule structure of the
Zeeman injection current in Eq.~\ref{eq_Fermi_Zeeman}.

The expressions for the Zeeman shift and injection currents (Eqs.~\ref{Zshift_purt} and~\ref{Zinj_purt}) constitute the central results of this work.
A finite interband $EB$ resonance requires spin mixing, which allows
electric-dipole and Zeeman transition amplitudes to interfere.
Such mixing naturally occurs in systems with spin--orbit coupling (SOC)
or noncollinear magnetic order, making them promising platforms
for realizing resonant $EB$ photocurrents.


\textbf{Symmetry Analysis}---
As shown in the phenomenological analysis, Zeeman shift and injection currents can remain finite in centrosymmetric systems. Using Eqs.~\ref{Zshift_purt} and~\ref{Zinj_purt}, we now examine their behavior under a broader set of discrete space-time symmetries and compare the resulting constraints with those for conventional electric-dipole shift and injection currents, as summarized in Table~\ref{tab:symmetry}. 
\begin{table}[t]
\centering
\caption{
Symmetry constraints on conventional and Zeeman shift and injection currents under spatial inversion $\mathcal{P}$, time reversal $\mathcal{T}$, and their combination $\mathcal{PT}$. $L$ ($C$) denotes a symmetry-allowed response to linearly (circularly) polarized light, whereas $\boldsymbol{\times}$ indicates a symmetry-forbidden photocurrent.
}
\label{tab:symmetry}
\begin{tabular*}{\columnwidth}{@{\extracolsep{\fill}}ccccc@{}}
\hline\hline
 & Shift & Injection & Zeeman shift &  Zeeman injection \\
\hline
$\mathcal{P}$
& $\boldsymbol{\times}$ & $\boldsymbol{\times}$ & $L/C$ & $L/C$ \\
$\mathcal{T}$
& $L$ & $C$ & $C$ & $L$ \\
$\mathcal{PT}$
& $C$ & $L$ & $C$ & $L$ \\
\hline\hline
\end{tabular*}
\end{table}


By analogy with the electric-dipole response, we decompose 
the $EB$ photoconductivity as $\sigma^{a;b\beta}=\sigma_L^{a;b\beta}+i\sigma_C^{a;b\beta}$,
where $\sigma_L$ and $\sigma_C$ denote its real and imaginary components, 
respectively. The $\sigma_L$ component is associated with
linearly polarized light, and $\sigma_C$ with circularly polarized light.


We first consider inversion symmetry $\mathcal{P}$, which highlights a key distinction between conventional and Zeeman photocurrents. Under inversion, $\mathcal{P}|n,\mathbf{k}\rangle=e^{-i\phi_n}|n,-\mathbf{k}\rangle$. The spin operator is inversion even, $\mathcal{P}\hat{\sigma}^{\beta}\mathcal{P}^{-1}=\hat{\sigma}^{\beta}$, whereas the electric-dipole operator is inversion odd. Their matrix elements therefore transform as $\sigma_{mn}^{\beta}(-\mathbf{k})=e^{i(\phi_n-\phi_m)}\sigma_{mn}^{\beta}(\mathbf{k})$ and $r_{nm}^{b}(-\mathbf{k})=-e^{i(\phi_m-\phi_n)}r_{nm}^{b}(\mathbf{k})$.

Consequently, the Zeeman geometry entering the $EB$ transition factor is inversion odd, $\mathcal{Z}_{nm}^{b\beta}(-\mathbf{k})=-\mathcal{Z}_{nm}^{b\beta}(\mathbf{k})$. The effective shift vector $\mathcal{R}_{mn}^{EB}=R_{mn}^{\beta;a}-R_{nm}^{b;a}$ is 
also inversion odd (see S87 and S88 in the Supplemental Material), as is the group-velocity difference  $\Delta_{mn}^{a}$.
The resulting Zeeman shift and injection kernels, $\mathcal{Z}_{nm}^{b\beta}\mathcal{R}_{mn}^{EB}$ and $\mathcal{Z}_{nm}^{b\beta}\Delta_{mn}^{a}$, 
are therefore inversion even, allowing both currents in centrosymmetric systems. This contrasts with conventional electric-dipole shift and injection currents, which are forbidden by inversion symmetry, as summarized in the first row of Table~\ref{tab:symmetry}.

We next examine time-reversal symmetry $\mathcal{T}$, using the Zeeman injection current as an example. Its linear and circular components satisfy
$\sigma_{L}(\mathbf{k})\propto\mathrm{Re}\mathcal{Z}_{nm}^{b\beta}\Delta_{mn}^a$
and
$\sigma_{C}(\mathbf{k})\propto\mathrm{Im}\mathcal{Z}_{nm}^{b\beta}\Delta_{mn}^a$, respectively, where 
$\mathrm{Re}\,Z_{nm}^{b\beta}
=(\mathcal{Z}_{nm}^{b\beta}+\mathcal{Z}_{mn}^{b\beta})/2$
and
$\mathrm{Im}\,\mathcal{Z}_{nm}^{b\beta}
=(\mathcal{Z}_{nm}^{b\beta}-\mathcal{Z}_{mn}^{b\beta})/(2i)$.
Under $\mathcal{T}$,
$\sigma_{mn}^{\beta}(\mathbf{k})
=-\sigma_{nm}^{\beta}(-\mathbf{k})$
and
$r_{mn}^{b}(\mathbf{k})
=r_{nm}^{b}(-\mathbf{k})$,
yielding
$\mathcal{Z}_{nm}^{b\beta}(\mathbf{k})
=-\mathcal{Z}_{mn}^{b\beta}(-\mathbf{k})$.
Consequently, $\mathrm{Re}\,\mathcal{Z}$ is $\mathcal{T}$-odd,
whereas $\mathrm{Im}\,\mathcal{Z}$ is $\mathcal{T}$-even.
Together with the $\mathcal{T}$-odd velocity difference, these properties render the linear Zeeman injection response $\mathcal{T}$-even and the circular response $\mathcal{T}$-odd. Thus, in $\mathcal{T}$-symmetric systems, Zeeman injection currents are allowed under linearly polarized light, reversing the polarization selection rule for conventional electric-dipole injection currents, which are driven by circularly polarized light. An analogous symmetry analysis applies to the Zeeman shift current (see Supplemental Material for details), with the resulting constraints summarized in the second row of Table~\ref{tab:symmetry}.

Finally, combined $\mathcal{PT}$ symmetry imposes similar constraints on Zeeman and conventional $EE$ photocurrents. We therefore present the detailed analysis in the Supplemental Material, with the results summarized in the third row of Table~\ref{tab:symmetry}.

Taken together, these symmetry results show that inversion symmetry forbids conventional $EE$ shift and injection currents irrespective of whether $\mathcal{T}$ is preserved or broken, allowing Zeeman ($EB$) photocurrents to be isolated.
Centrosymmetric materials with strong SOC, such as bulk Bi$_2$Se$_3$, and inversion-symmetric noncollinear magnets thus provide promising platforms for investigating these effects.

\begin{figure}[!t]
\includegraphics[width=8.5 cm]{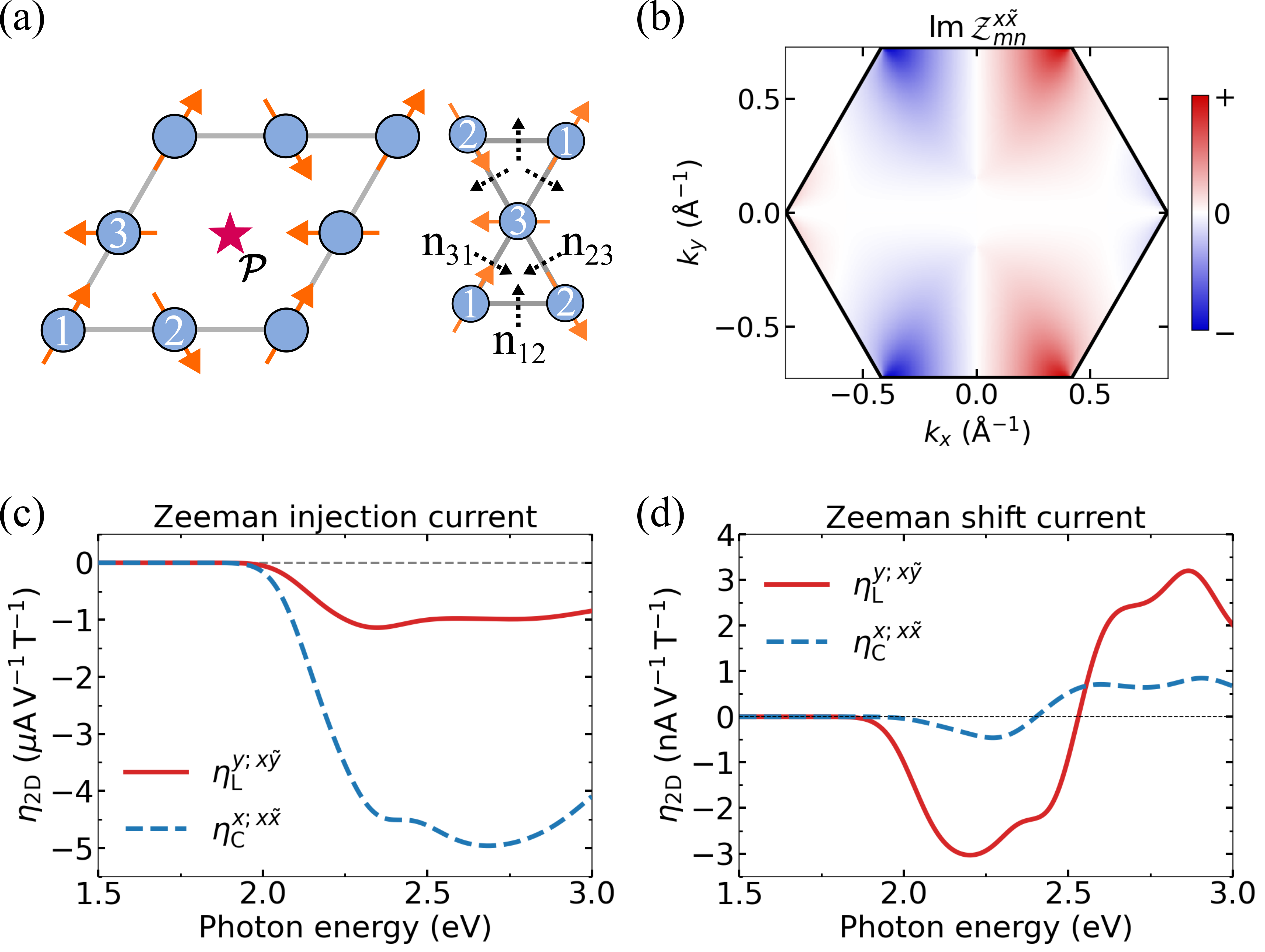}
\caption{
(a) Schematic of the kagome model with coplanar magnetic order. Orange and black arrows denote local magnetic moments and bond-dependent SOC, respectively. The red star marks the inversion center $\mathcal{P}$.
(b) Momentum-space distribution of the $EB$-transition factor
$\mathrm{Im}\,\mathcal{Z}_{mn}^{x\tilde{x}}$ between bands $m=6$ and $n=1$.
(c) Photon-energy dependence of the Zeeman injection responses to linearly and circularly polarized light, $\eta_{L}^{y;x\tilde{y}}$ and $\eta_{C}^{x;x\tilde{x}}$, respectively.
(d) Corresponding Zeeman shift responses.
The model parameters are $t=1~\mathrm{eV}$,
$J=1.7~\mathrm{eV}$, $\lambda_{\mathrm{SO}}=0.2~\mathrm{eV}$ and $\tau = 1\ \mathrm{ps}$.
}
\label{Fig2}
\end{figure}

\textbf{Zeeman photocurrents in a centrosymmetric kagome model}---
To illustrate these mechanisms, we consider a kagome model with $120^\circ$ coplanar antiferromagnetic order.
Following Chen et al.~\cite{Chen2014}, we write the Hamiltonian as
\begin{align}
H=&t\sum_{\langle i\alpha,j\beta\rangle\gamma}
c_{i\alpha\gamma}^\dagger c_{j\beta\gamma}
-J\sum_{i\alpha\gamma\tau}
c_{i\alpha\gamma}^\dagger
(\mathbf{m}_\alpha\cdot\boldsymbol{\sigma}_{\gamma\tau})
c_{i\alpha\tau}
\notag\\
&+\mathrm{i}\lambda_{\rm SO}
\sum_{\langle i\alpha,j\beta\rangle\gamma\tau}
c_{i\alpha\gamma}^{\dagger}
(\mathbf{n}_{\alpha\beta}\cdot
\boldsymbol{\sigma}_{\gamma\tau})
c_{j\beta\tau}.
\end{align}
where $i,j$ label the unit cells, 
$\alpha,\beta=1,2,3$ the kagome sublattices, 
and $\gamma,\tau=\uparrow,\downarrow$ the spin indices.
The first term describes nearest-neighbor hopping with amplitude $t$, and the second represents 
exchange coupling between the electron 
spin and the magnetic texture $\mathbf{m}_\alpha$, 
indicated by the orange arrows 
in Fig.~\ref{Fig2}(a).
The third term describes bond-dependent SOC, characterized by the vectors $\mathbf{n}_{\alpha\beta}=-\mathbf{n}_{\beta\alpha}$ shown as black dashed arrows in Fig.~\ref{Fig2}(a).

The red star in Fig.~\ref{Fig2}(a) marks the model’s inversion center.
We next examine the inversion parity of the $EB$ transition factor, taking $\mathrm{Im}\,\mathcal{Z}_{mn}^{x\tilde{x}}$ for bands $m=6$ and $n=1$ as an example. Here, $x$ and $\tilde{x}$ label the electric- and magnetic-field directions, respectively; tildes denote magnetic-field indices hereafter. As shown in Fig.~\ref{Fig2}(b), this factor changes sign under $\mathbf{k}\rightarrow-\mathbf{k}$, reflecting its $\mathcal{P}$-odd character.
Since the group-velocity difference and shift vector are also $\mathcal{P}$-odd, the Zeeman injection- and shift-current integrands are $\mathcal{P}$-even in momentum space. Consequently, Zeeman photocurrents can remain finite in centrosymmetric systems, whereas conventional $EE$ shift and injection currents are forbidden.

Figs.~\ref{Fig2}(c) and (d) show the calculated two-dimensional Zeeman injection and shift photoconductivities, respectively, including the linear component $\eta_{L}^{y;x\tilde{y}}$ and the circular component $\eta_{C}^{x;x\tilde{x}}$.
Both Zeeman shift and injection currents exhibit finite over a broad photon-energy range, providing a direct microscopic demonstration of symmetry-allowed Zeeman photocurrents in a centrosymmetric system.

\begin{figure}[t]
\includegraphics[width=8.5cm]{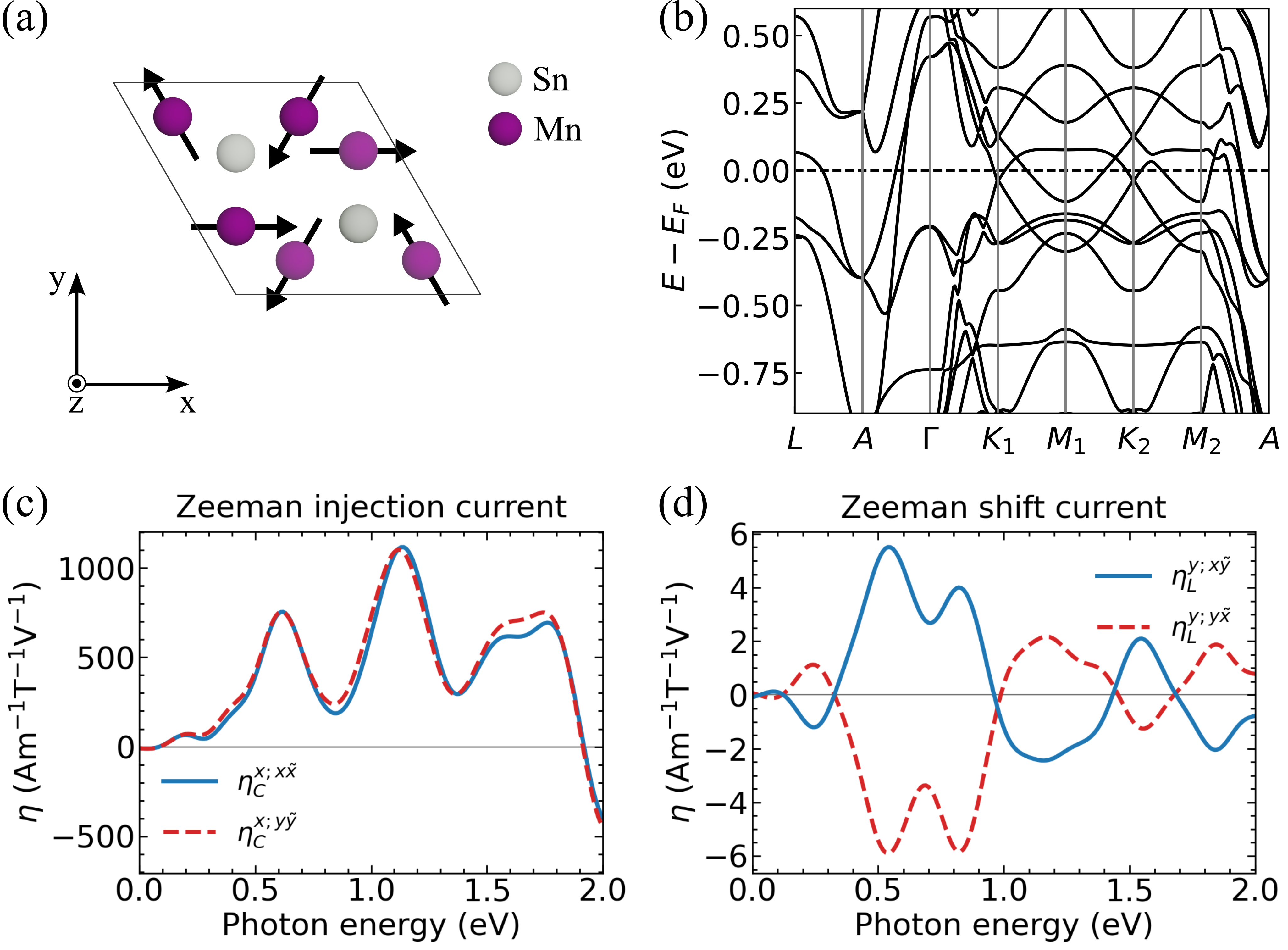} 
\caption{ 
(a) Crystal structure of $\mathrm{Mn_3Sn}$, with black arrows indicating the coplanar magnetic order.
(b) Electronic band structure.
(c) Circularly polarized components of the Zeeman injection current,
$\eta_{C}^{x;x\tilde{x}}$ and $\eta_{C}^{x;y\tilde{y}}$, as functions
of photon energy.
(d) Linearly polarized components of the Zeeman shift current,
$\eta_{L}^{y;x\tilde{y}}$ and $\eta_{L}^{y;y\tilde{x}}$, as functions
of photon energy.
} 
\label{Fig3}
\end{figure}

\textbf{Observable Zeeman Photocurrents in $\mathbf{Mn_3Sn}$}---
Having established Zeeman photocurrents in the kagome model, we turn to the noncollinear antiferromagnet Mn$_3$Sn as a realistic material platform~\cite{Nakatsuji2015,Higo2018}.
Its coplanar magnetic structure shown in Fig.~\ref{Fig3}(a) belongs to the centrosymmetric magnetic space group $Cmc'm'$~\cite{Tomiyoshi1982,Cederholm2026}. The resulting absence of conventional $EE$ shift and injection currents makes Mn$_3$Sn a particularly clean platform for observing Zeeman photocurrents.

The calculated band structure of Mn$_3$Sn in Fig.~\ref{Fig3}(b) is consistent with previous first-principles results~\cite{Kuroda2017} (see Supplemental Material~\cite{supp,Kresse1996,perdew1996} for computational details).
We consider light propagating along $z$, with its electric and magnetic fields in the $xy$ plane. In this geometry, the magnetic symmetry of Mn$_3$Sn allows an in-plane Zeeman injection current only under circularly polarized light. The two allowed components, $\eta_{C}^{x;x\tilde{x}}$ and $\eta_{C}^{x;y\tilde{y}}$, are shown in Fig.~\ref{Fig3}(c). 
With SOC included, no symmetry enforces equality between these two components, consistent with the small difference observed here.
The three-dimensional Zeeman photoconductivity reaches magnitudes of order $10^3~\mathrm{A/(mTV)}$ over a broad photon-energy range for a representative relaxation time of $\tau=1~\mathrm{ps}$.
In contrast, the in-plane Zeeman shift current is allowed only under linearly polarized light. As shown in Fig.~\ref{Fig3}(d), the nonzero components $\eta_{L}^{y;x\tilde{y}}$ and $\eta_{L}^{y;y\tilde{x}}$ reach magnitudes of several $\mathrm{A/(mTV)}$. All calculations use a conservative value of $g=2$ for the electron spin $g$ factor.


To assess experimental feasibility, we consider a representative optical electric-field amplitude $E=10^4~\mathrm{V/m}$, corresponding to $B=E/c\simeq3\times10^{-5}~\mathrm{T}$ for propagating light.
The estimated Zeeman injection and shift current densities reach $10^{2}$--$10^{3}~\mathrm{A/m^{2}}$ and \(1\)–\(10~\mathrm{A/m^{2}}\), respectively. For comparison, conventional shift-current densities in noncentrosymmetric materials typically reach \(10^{2}\)–\(10^{3}~\mathrm{A/m^{2}}\) at the same electric-field amplitude. The reported observation of conventional shift currents under weak illumination~\cite{Choi2009,Grinberg2013,Hatada2020,Nakamura2017}, including simulated sunlight, suggests that both Zeeman responses should be experimentally accessible. Owing to its larger magnitude, the Zeeman injection current should be more readily detectable than the Zeeman shift current.

In summary, we have established Zeeman shift and injection currents arising from interference between the electric-dipole and spin Zeeman couplings of light. These $EB$ responses are allowed in systems with diverse symmetries. Their distinct inversion parity enables their isolation in centrosymmetric materials, where conventional $EE$ shift and injection currents vanish. Our microscopic theory and symmetry classification guide the identification of suitable materials and polarization geometries for observing these effects.
More broadly, these findings reveal how the magnetic component of light couples electronic spin textures to charge transport, providing a route to accessing spin information through dc photocurrents. This connection suggests opportunities for electrical detection of magnetic states and spin-sensitive photodetection.

\textbf{Acknowledgments-}
This work was supported by the National Natural Science 
Foundation of China (Grant No.~12204035)
and the National Key R\&D Program of China (Grant No.~2025YFA1411200).



\nocite{showArticleTitles}
\bibliographystyle{apsrev4-2}
\bibliography{optical}

\end{document}